\documentclass[journal]{IEEEtran}
\usepackage{array,multirow}
\usepackage{graphicx}
\usepackage{caption}
    \DeclareCaptionFont{mysize}{\fontsize{8}{8}\selectfont}
\usepackage{tabularx}
\usepackage{tabulary}
\usepackage{latexsym}
\usepackage{textgreek}
\usepackage{layout}
\usepackage{rotating}
\usepackage{amsmath,amsfonts,amssymb}
\usepackage{bm}
\usepackage{euscript}
\usepackage{algorithmic}
\usepackage{amssymb, amsmath}
\usepackage[numbers, square, comma, sort&compress]{natbib}
\usepackage[ruled,lined]{algorithm2e}
\usepackage{epstopdf}
\usepackage{gensymb}
\usepackage{subcaption}
\usepackage{siunitx}
\usepackage{url}
\usepackage{xcolor}
\usepackage{ragged2e}
\usepackage[english]{babel}
\usepackage{multicol}
\usepackage[official]{eurosym}
\usepackage{color}
\usepackage{acronym}
\usepackage{mathtools}
\usepackage[verbose]{wrapfig}
\usepackage{multirow}
\usepackage[normalem]{ulem}

\usepackage [autostyle, english = american]{csquotes}
\MakeOuterQuote{"}
\usepackage{caption} 
   \usepackage[belowskip=2pt, aboveskip=-13pt] {caption}
\acrodef{API}[API]{Application Programmable Interfaces}
\acrodef{BS}[BS]{Base Station}
\acrodef{CNN}[CNN]{Convolutional Neural Network}
\acrodef{CPU}[CPU]{Central Processing Unit}
\acrodef{DETA-R}[DETA-R]{Dynamic and Energy-Traffic-Aware algorithm with Random behavior}
\acrodef{EB}[EB]{Energy Buffer}
\acrodef{EH}[EH]{Energy Harvesting}
\acrodef{ES}[ES]{Energy Saving}
\acrodef{EM}[EM]{Energy Manager}
\acrodef{ENAAM}[ENAAM]{Energy Aware and Adaptive Management}
\acrodef{GENM}[GENM]{Green-based Edge Network Management}
\acrodef{EPC}[EPC]{Evolved Packet Core}
\acrodef{ITS}[ITS] {Intelligent Transport System}
\acrodef{LOC}[LOC]{User Location Services} 
\acrodef{LLC}[LLC]{Limited Lookahead Control}
\acrodef{LS}[LS]{Location service}
\acrodef{LSTM}[LSTM]{Long Short-Term Memory}
\acrodef{MEC}[MEC]{Multi-access Edge Computing}
\acrodef{ML}[ML]{Machine Learning}
\acrodef{MN}[MN]{Mobile Network}
\acrodef{TIM}[TIM]{Telecom Italia Mobile}

\acrodef{NOES}[NOES]{NO Energy Saving}
\acrodef{NFV}[NFV]{Network Function Virtualization}
\acrodef{NIC}[NIC]{Network Interface Card}
\acrodef{QoS}[QoS]{Quality of Service}
\acrodef{RNN}[RNN]{Recurrent Neural Network}
\acrodef{RAN}[RAN]{Radio Access Network}
\acrodef{RMSE}[RMSE]{Root Mean Square Error}
\acrodef{RNN}[RNN]{Recurrent Neural Network}
\acrodef{UE}[UE]{User Equipment}
\acrodef{VLAN}[VLAN]{Virtual Local Area Network}
\acrodef{VM}[VM] {Virtual Machines}
\acrodef{VNF}[VNF]{Virtualized Network Function}

\begin{document}

\title{Core Network Management Procedures for Self-Organized and Sustainable 5G Cellular Networks}

\author{\IEEEauthorblockN {Thembelihle Dlamini\IEEEauthorrefmark{1}}\\
	\IEEEauthorblockA {\IEEEauthorrefmark{1}Department of Information Engineering, University of Padova, Padova, Italy}\\
	dlamini@dei.unipd.it \vspace{-0.4cm}
}

\maketitle
\thispagestyle{plain}
\pagestyle{plain}

\begin{abstract}

Future Mobile Networks (MNs), 5G and beyond 5G, will require a paradigm shift from traditional resource allocation mechanisms as Base Stations (BSs) will be empowered with computation capabilities (i.e., offloading and computation is performed closer to mobile users). This is motivated by the expected data explosion in the volume, variety, and velocity, generated by pervasive mobile and Internet of Things (IoT) devices at the network edge. Towards efficient resource management, within the Multi-access Edge Computing (MEC) paradigm, we make use of the Long \mbox{Short-Term} Memory (LSTM) neural network for time series forecasting and control-theoretic techniques for foresighted optimization, thus bringing intelligent mechanisms for handling network resources within the network edge. Here, we propose online algorithms for autoscaling and reconfiguring the computing-plus-communication resources within the virtualized computing platform, and also enable dynamic switching on/off BSs by taking into account the forecasted traffic load and harvested energy. The main goal is to minimize the overall energy consumption, with a guarantee of Quality of Service (QoS). Our numerical results, obtained through trace-driven simulations, show that the proposed optimization strategies lead to a considerable reduction in the energy consumed by the edge computing and communication facilities, promoting energy self-sustainability within the MN through the use of green energy.
\end{abstract}


\IEEEpeerreviewmaketitle

\section{Introduction}

The data growth generated by pervasive mobile devices and the Internet of Things, coupled with the demand for \mbox{ultra-low} latency, requires high computation resources which are not available at the \mbox{end-user} device. As a remedy, {\it\ac{MEC}} has recently emerged to enable \mbox{ultra-low} latency, distributed intelligence and \mbox{location-aware} data processing at the network edge~\cite{etsimec_access}\cite{etsimec}. Undoubtedly, offloading to a powerful computational \mbox{resource-enriched} server located closer to mobile users is an ideal solution. Specifically, by offloading data- and computing-intensive tasks to the nearby \ac{MEC}-enabled \acp{BS}, resource-constrained mobile devices benefit from reduced energy consumption and task completion time. Although this new approach is not intended to replace the \mbox{cloud-based} infrastructure, it expands the cloud by increasing the computing and storage resources available at the network edge. 

In light of the dense deployment pattern that is foreseen in $5$G systems~\cite{bhushan2014network}, the expected dense deployment of \ac{MEC} servers and \acp{BS} raises concerns related to \textit{energy consumption}. Specifically, energy drained in \acp{BS} is due to the \mbox{always-on} approach and in \ac{MEC} servers it is due to the computing and communication processes associated with: (i) the running \acp{VM}~\cite{virttech}\cite{eempirical}; (ii) the communication within the server's \ac{VLAN}~\cite{vm_book}, and the presence of transmission drivers ({\it fast} tunable optical drivers for data transfers) within the \ac{MN} infrastructure~\cite{laser_tuning}. Towards greener \acp{MN}, there is a need for efficient edge network management procedures that will yield the \mbox{trade-off} between energy savings and the guarantee of \ac{QoS}.
Thus, given the expected traffic load and harvested energy, the energy consumption can be minimized by launching an optimal number of \acp{VM}, a technique referred to as VM \mbox{\it soft-scaling}, tuning of the transmission drivers {\it coupled} with the \mbox{location-aware} traffic routing for data transfer, together with BS power savings, i.e., enabling BS sleep modes.

Along these lines, we consider an energy cost model that takes into account the computing and communication processes within the \ac{MEC} server, and  transmission related energy in \acp{BS},  where the edge network apparatuses are energized by green energy. We propose an online edge system management procedure/algorithm, called \ac{ENAAM}, that dynamically allows the switching on/off BSs, \acp{VM}, transmission drivers (i.e., using a minimum number of drivers for \mbox{real-time} data transfers), depending on the traffic load and the harvested energy forecast, over a given lookahead prediction horizon. To solve the energy consumption minimization problem, we leverage the \ac{LLC} principles where the controller performs online supervisory control by forecasting the traffic load and the harvested energy using a \ac{LSTM} neural network~\cite{lstmlearn}, which is utilized within a \ac{LLC} policy (a predictive control approach~\cite{llcprediction}) to obtain the system control actions that yields the desired trade-off between energy consumption and \ac{QoS}. 

The proposed optimization strategy leads to a considerable reduction in the energy consumed by the edge computing and communication facilities, promoting self-sustainability within the \ac{MN} through the use of green energy. This is achieved under the controller guidance, which makes use of forecasting, \ac{LLC} principles and heuristics.

\begin{figure*} [h!]
   \begin{minipage}{0.5\linewidth}
      \resizebox{\columnwidth}{!}{\input{load_pred_info.tex}}
	  \caption{One-step ahead predictive mean value for $L(t)$ \\and real values.}
	  \label{fig:bs_load_2}	
   \end{minipage}%
   \begin{minipage}{0.5\linewidth}
      \resizebox{\columnwidth}{!}{\input{energy_pred_info.tex}}
	  \caption{One-step ahead predictive mean value for $H(t)$ \\and real values.}
	  \label{fig:energy_load_2}	
   \end{minipage}
\end{figure*}

\section{Methodology}

Let $\textupsilon(t)$ be the current system vector and $\psi(t)$ is the input vector that drives the system into the desired behavior, at time $t$. This work adopts foresighted optimization and the system behavior $\textupsilon(t + 1)$ is described by the \mbox{discrete-time} \mbox{state-space} equation, adopting the \ac{LLC} principles~\cite{llcprediction}:
\begin{equation}
\textupsilon(t + 1) = \phi(\textupsilon(t), \psi(t)) \, , 
\end{equation}
\noindent where  $\phi(\cdot)$ is a behavioral model that captures the relationship between $(\textupsilon(t),\psi(t))$, and the next state $\textupsilon(t + 1)$. Since the actual values for the system input cannot be measured until the next time instant when the controller adjusts the system behavior, the corresponding system state for $t+1$ can only be estimated as:
\begin{equation}
       \hat{\textupsilon}(t + 1) = \phi(\textupsilon(t),\psi(t)) \,.
       \label{eq:state_forecast}
\end{equation}
For these estimations we use the forecast values of load and harvested energy, from the LSTM forecasting module.

\section{Numerical Results}


Figs.~\ref{fig:bs_load_2} and~\ref{fig:energy_load_2}, show real and predicted values for the traffic load $L(t)$ and harvested energy $H(t)$ over time, where we track the \mbox{one-step} predictive mean value at each step. The average \ac{RMSE} are: $L(t) =\{ 0.037, 0.042, 0.048\}$ and $H(t) = \{0.011, 0.016, 0.021\}$, for time horizon $T = 1,2,3$. Note that the predictions for $H(t)$ are more accurate than those of $L(t)$ (see RMSE values), due to differences in the used dataset granularity.

\begin{figure}
\resizebox{\columnwidth}{!}{\input{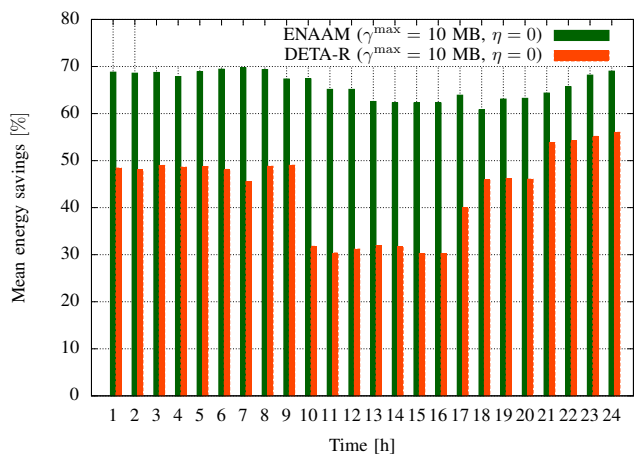}}
\caption{Mean energy savings for $\eta = 0$ and $\gamma^{\max} = 10$ MB.}\label{fig:savings_010}	
\end{figure}

Fig.~\ref{fig:savings_010}, shows the mean energy savings of $69\%$ for the \ac{ENAAM} scheme, which makes use of LSTM and LLC (maximum VM load, $\gamma^{\max} = 10$~MB), while \ac{DETA-R} achieves $49\%$ ($\gamma^{\max} = 10$~MB), where these savings are with respect to the case where {\it no energy management} is performed. The mean values were obtained at weighing parameter $\eta = 0$.

\section*{Acknowledgements}

This work has received funding from the European Union's Horizon 2020 research and innovation programme under the Marie Sklodowska-Curie grant agreement No. 675891 \mbox{(SCAVENGE)}.

\bibliographystyle{IEEEtran}
\scriptsize
\bibliography{biblio_new}
\end{document}